\documentclass[prl,aps,amsmath,amssymb,reprint]{revtex4-1}
\usepackage{graphicx}
\usepackage{epstopdf}
\usepackage{dcolumn}
\usepackage{bm}
\newcommand{\p}{$\%$}

\newcommand{\pn}{$\mathrm{R{_{N_2}}}$}

\newcommand{\tcn}{$\mathrm{Co_{4}N}$}

\newcommand{\muB}{$\mathrm{\mu_{B}}$}

\newcommand{\Ts}{$\mathrm{T_{s}}$}

\newcommand{\mB}{$\mu\mathrm{_{B}}$}
\newcommand{\dE}{$\Delta{H{_{f}^{\circ}}}$}
\newcommand{\ms}{$\textbf{\emph{M}}$}

\begin{document}
\title{Structural, Electronic, and Magnetic Properties of HiPIMS Grown Co-N Thin Films}
\author{Seema$^1$, Akhil Tayal$^2$, S. M. Amir$^3$, Sabine P\"{u}tter$^3$, S. Mattauch$^3$, and Mukul Gupta$^1$}
\email {mgupta@csr.res.in}
\affiliation{$^1$UGC-DAE Consortium for Scientific Research, University Campus, Khandwa
Road, Indore 452 001, India}
\affiliation{$^2$Deutsches Elektronen-Synchrotron DESY, Notkestrasse 85, D-22607 Hamburg, Germany}
\affiliation{$^3$J\"{u}lich Centre for Neutron Science (JCNS) at Heinz Maier-Leibnitz Zentrum (MLZ) Forschungszentrum J\"{u}lich GmbH, Lichtenbergstr. 1, 85747 Garching, Germany}

\date{\today}


\begin{abstract}
We studied the growth behavior, structural, electronic, and magnetic
properties of cobalt nitride (Co-N) thin films deposited using direct current (dc) and high
power impulse magnetron sputtering (HiPIMS) processes. The N$_2$
partial gas flow (\pn) was varied in close intervals to achieve
the optimum conditions for the growth of tetra cobalt nitride
(\tcn) phase. We found that
Co-N films grown using HiPIMS process adopt (111) orientation as compared
to the growth taking place along the (100) direction in the dcMS
process. It was observed that HiPIMS grown Co-N~films were superior in terms of crystallite size and uniform surface morphology. The local structure of films was investigated using x-ray absorption fine structure (XAFS) measurements. We found that the high energy of adatoms in the HiPIMS technique assisted in the greater stabilization of fcc-Co and novel \tcn~phase relative to the dcMS process. Magnetic properties of Co-N thin films were studied using magneto-optical Kerr effect, vibrating sample magnetometry and polarized neutron reflectivity. It was found that though the saturation magnetization remains almost similar in films grown by dcMS or HiPIMS processes, they differ in terms of their magnetic anisotropy. Such variation can be understood in terms of differences in the growth mechanisms in dcMS and HiPIMS processes affecting the local structure of resulting \tcn~phase.
\end{abstract}

\maketitle

\section{Introduction}
\label{1} Transition metal nitrides (TMNs) are a class of interstitial
compounds that can be placed somewhere in between pure metals and
metal-oxides. Generally, TMNs are known to possess metal like resistivity and the hybridization between metal-3d and N-2p orbitals leads to several interesting outcomes; e.g. excellent magnetic, electrical, and chemical
properties~\cite{1993_PRB_Haglund}. TMNs appearing late in 3$d$ series are emerging compounds having wide range of applications from spintronic devices to electrocatalysts in hydrogen and oxygen evolution reactions~\cite{meng2018}. Among late TMNs, Co-N system has been relatively less explored. Co forms nitride compounds in a wide structural and compositional range as N content is
increased~\cite{APL:86:CoN}. Among various Co-N phases, the \tcn~phase has received the most attention.~\cite{1987_JMS_Oda,2015_Silva,JVST99_Zhang,2011_JCG_Ito_Co4N_STO,2014_JAP_Ito_CoN,PRB:CoN:07} This is because \tcn~exhibits several interesting properties such as: (i) high saturation magnetization (\ms) $\approx$ 1.61\,\mB/Co atom~\cite{PRB:CoN:07} (ii) high spin-polarization ratio (SPR) $\approx$~90\p~\cite{PRB:CoN:07} (iii)
a half-metallic nature (iv) corrosion resistance (v) ultrahigh
electrical conductivity~\cite{chen2016cobalt} and (vi) a high Curie
temperature of about 1027\,K.~\cite{JPCM_16_Markus} Superior
magnetic properties make \tcn~as a promising ferromagnetic
material to be used in spintronics and magnetic recording
media, while the catalytic properties of
\tcn~make its usage as a reducing agent in oxygen evolution reactions and electrode material in metal-air batteries~\cite{1987_JMS_Oda,chen2016cobalt,fan2019ALD}.

Despite excellent chemical and magnetic properties, the phase
diagram for Co-N system has not been established yet. Over past few decades Co-N compounds have been synthesized by nitridation of Co metal~\cite{zhong2015},
thermal decomposition and ammonolysis of Co based
compounds~\cite{niewa2016Co-N}, atomic layer
deposition (ALD)~\cite{fan2019ALD}, metal organic chemical vapor
deposition (MOCVD)~\cite{2003_Co_K_YO}, molecular beam epitaxy
(MBE)~\cite{2014_JAP_Ito_CoN}, reactive dc/rf magnetron
sputtering(R-MS).~\cite{JVST99_Zhang,fang2004phase,2015_Silva,JAC16_NPandey}
Among these, samples prepared using R-MS and MBE were claimed
to be single phase \tcn~\cite{2014_JAP_Ito_CoN,2015_Silva}. However, even in these studies, the lattice parameter (LP) of \tcn~thin film sample was always found to be
significantly lower (3.54 to 3.67\AA) as compared to its
theoretically predicted value of 3.73~\AA. As the enthalpy of formation of \tcn~is $\approxeq$~0\,eV, samples have been prepared at high substrate temperature (\Ts).~\cite{imai2014} It has been proposed that N out-diffusion from \tcn~at high \Ts~leads to expanded fcc-Co lattice formation (LP=3.58\,\AA), that seems to be misunderstood as \tcn.~\cite{JAC16_NPandey} Another possibility can be that in the case of thin films deposition processes, several parameters influence the structure of the compound. Thermodynamical conditions in the nanoparticle systems can stabilize the structure under ambient conditions, that are otherwise met under extremely high pressure and/or high temperature.~\cite{zhang2017materials} 

The inherent nature of the most widely used sputtering technique induces defects and disorder that could also lead to the formation of random stacking sequences.~\cite{JAP_Co_C,MRSB_2016} The cumulative effect of these parameters and significantly small stacking fault energy cause the formation of heterogeneous structures, that has been overlooked in the synthesis of cobalt nitrides. Therefore, it is highly possible that the films can have a mixture of hcp-Co, fcc-Co, together with a non-stoichiometric \tcn~phase. This makes the precise determination of the structure of Co or closely related Co-N phases an exceptional challenge. Since, the structure of \tcn~and fcc-Co is similar and the only distinction between them is that in \tcn~all the octahedral interstitial sites are occupied with the nitrogen. Filling of octahedral sites also results in an increment in the LP, leading to an expanded lattice of fcc-\tcn~as compared to fcc-Co. Therefore, the distinction between fcc-Co and \tcn~become very difficult specially when the octahedral sites are partially occupied.

From the above description, it obvious that the formation of a single phase \tcn~is challenging and a critical inspection of the structure is required. Also, the \tcn~phase needs to
be synthesized without raising the \Ts~to prevent N diffusion from \tcn.~\cite{PhysRevB.NP} Some attempts have been made
using MBE and R-MS to grow \tcn~phase at low \Ts. While using the MBE process, the \tcn~phase does not form at all at low
\Ts~\cite{JMMM_MG}, but the signatures of \tcn~phase can be observed in samples deposited using R-MS process at
\Ts\ = 300\,K.~\cite{CoN_AIP_Adv2015,2001_Vac_Asahara,JMMM_MG} However, the \tcn~formed with R-MS at 300\,K does not seem to be single phase due to presence of some impurity phase.~\cite{JMMM_MG} Therefore, the absence of substrate heating makes the formation of stoichiometric \tcn~phase even more challenging.

In view of this, an emerging technique known as high power impulse magnetron sputtering (HiPIMS) is a promising deposition method due to the tunability of adatom
energies.~\cite{SARAKINOS2010,ANDERS2011} This technique has been
characterized as ionized physical vapor deposition (IPVD) which
enhance film qualities such as microstructure, density, hardness,
adhesion and, reduced surface
roughness.~\cite{HiPIMS:Hala:CrN:JAP10} In this process, the high
ion flux enhances surface mobility of the adatom and thereby
reducing film porosity.~\cite{HiPIMS:Gudmundsson} This enhanced adatom mobility in HiPIMS plasma can work as a substitute for the need of high \Ts. 

In the present
work, we optimized growth conditions for \tcn~phase by varying
nitrogen partial pressure (\pn) in close intervals using HiPIMS and
compared these films with those grown using dcMS in the same deposition chamber at ambient condition. A comparison for structural and magnetic properties have been made. We have utilized x-ray absorption fine structure (XAFS) technique, which provides information of the local structure and is sensitive to all the structural motif irrespective of their degree of crystallization. This analysis was performed to investigate the evolution of local structure in the samples deposited at different \pn~using dcMS and HiPIMS process. Therefore, the present study is expected to unravel any correlation between observed distinct microstructural and magnetic properties of Co-N thin films deposited using dcMS and HiPIMS processes.

\section{Experimental Details}
\label{2} Polycrystalline Co-N thin films were deposited on amorphous quartz
(SiO$_2$) using dcMS and HiPIMS process.~\cite{JMMM_MG} The partial pressuere of N$_2$ gas defined as \pn~(\p)=
100$\times$p$_{\mathrm{N}_2}$/(p$_{\mathrm{Ar}}$+p$_{\mathrm{N}_2}$);
where, p$_{\mathrm{Ar}}$ and p$_{\mathrm{N}_2}$ are the gas flow of Ar and
N$_2$ gases, was vaired in small steps of 2.5\p, so as to achieve the optimum stoichiometric conditions for \tcn~phase.~\cite{CoN_AIP_Adv2015,2018_JMMM_NP} A base pressure of 2$\times$10$^{-8}$\,Torr was always achieved prior to deposition of a Co-N sample. In the dcMS process, the target power was kept fixed at 180\,W and the working pressure was kept at 3\,mTorr. In the HiPIMS process, the deposition parameters used were: average target power 200\,W; peak power 4.4\,kW (peak current of 7.4\,A), working pressure 3\,mTorr. The applied voltage to the target was $\approx$~600\,V. The power pulse lasted 90\,$\mu$s, and the pulse frequency
was 500\,Hz, yielding a duty cycle (DC) of 4.5\p, here the DC is defined as: DC=${t_\mathrm{ON}}/({t_\mathrm{ON}+t_\mathrm{OFF}})$. This value of the DC was chosen considering the optimized conditions achieved for the growth of TiN phase separately following the reports for HiPIMS process by where a moderate DC~$\approxeq$~5\p~leads to dense and stoichiometric TiN films.~\cite{lattemann2010,chang2014} Typical thickness of all Co-N films was about 100\,nm. Deposition
rates and density for Co and \tcn~phase deposited using dcMS and HiPIMS were calculated from x-ray
reflectivity and discussed elsewhere.~\cite{JMMM_MG}

After deposition, the structural
characterization was done by x-ray diffraction (XRD) using a
standard x-ray diffractometer (Bruker-D8 Advance) using Cu
K$_\alpha$ x-rays. In order to get a detailed information about the atomic environment, XAFS measurements were performed in the fluorescence geometry at the P64 beamline at DESY, Hamburg, Germany~\cite{P64_ref}. A passivated implanted planar silicon detector was used to measure the integrated fluorescence signal. Energy scans around Co K-edge (7709\,eV) were performed using Si(111) monochromator. For the normalization of the XAFS spectra a Victoreen polynomial function is fitted in the pre-edge region and extrapolated till the end energy of a scan. This function was then subtracted from the data to remove the pre-edge background.~\cite{Teo:1981} The extended x-ray absorption fine structure (EXAFS) signal was extracted using Athena software and Autoback algorithm~\cite{IFEFFIT:Ravel:2005}. Fourier transform (FT) of data was performed in the k-range between 3-16~\AA$^{-1}$~using a sine window with $k^2$~weighting. The processing of the data was also performed in the Athena software~\cite{IFEFFIT:Ravel:2005} and the EXAFS fitting was performed in the k-space using a software code written by Conradson $et~al.$~\cite{Conradson:2013} The photoelectron backscattering amplitude and phase shifts were theoretically calculated using FEFF9 code~\cite{Feff9} using crystallographic structures of hcp-Co, fcc-Co and \tcn~phases. The simulated spectra have been included in the supplemental material (SM~\cite{SM}), which shows the variation in structure of hcp-Co, fcc-Co and \tcn. Additionally, the FT spectra for Co$_3$N had also been simulated to check the formation of Co$_3$N phase at high \pn.

Magnetic properties were studied using PNR, MOKE
(Evico Magnetics) equipment in longitudinal~(L) and polar~(P)
geometry and S-VSM (Quantum Design) at room temperature. PNR measurements were performed at MAgnetic Reflectometer with high
Incident Angle (MARIA), at JCNS, Garching, Germany in the
horizontal scattering plane geometry.~\cite{2018_MARIA} During PNR
measurements, samples were saturated by applying a
magnetic field of 0.5\,T parallel to the sample surface at room temperature and the data
was fitted using GenX software.~\cite{GenX}

\section{Results and Discussion}
\label{3}
\subsection{\tcn~phase formation using dcMS}\label{3.1}

\begin{figure}\center
	\includegraphics [width=85mm] {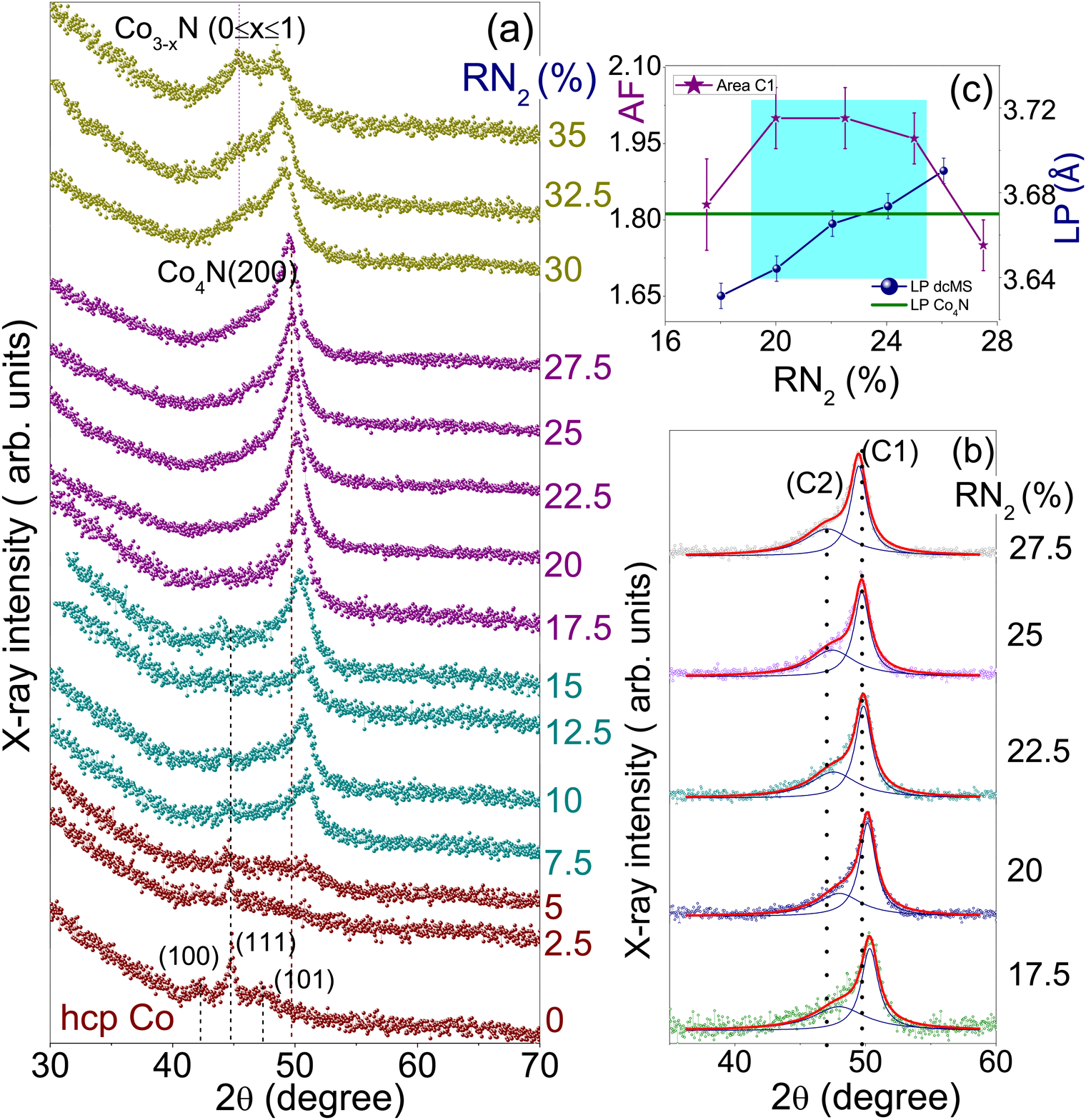}
	\caption{\label{fig:xrddcms} (a) XRD patterns of Co-N thin films deposited at different \pn using the dcMS process. (b) Fitted XRD data of selected samples assuming two Gaussian
		components C1 (\tcn~phase) and C2 (impurity phase). (c) The variation of area fraction (AF) of
		the component C1 and obtained values of lattice parameter (LP) at different \pn.}
	\vspace{-1mm}
\end{figure}

XRD data of Co-N thin film samples prepared using the dcMS process are shown in
fig.~\ref{fig:xrddcms}(a). In pure Ar sputtered film (\pn~=~0\p)
peaks can be seen at 2$\theta$ = 42.22, 44.82 and 47.23$^{\circ}$
($\pm$0.05$^{\circ}$) corresponding to hcp Co [JCPDS\#89-7094].
When a small amount of nitrogen is introduced in the process
chamber, the hcp structure distorts and the peak intensity starts to
diminish. This can be clearly seen as the variation in diffraction pattern for
\pn~=~2.5 and 5\p. Such a loss in the long range crystalline
ordering upon reactive nitrogen sputtering with a small amount of
nitrogen is a general phenomenon that can be seen in case of
several metals like Fe, FeNi~\cite{gupta2005}, Cr~\cite{YNT2019} etc.

When \pn~is increased to 7.5\p~and beyond, the structure
transforms from hcp to fcc. The peak appearing at 2$\theta$
$\approx$~50.38$^{\circ}$ can be assigned to (200) reflection of
fcc \tcn.~\cite{CoN_AIP_Adv2015,jia2008} Further N insertion keeps
on shifting this peak to lower 2$\theta$, which means that the fcc
Co lattice keeps on expanding as a function of N incorporation in
the host Co. From \pn~=~7.5 to 30\p, the LP increases from 3.578 to 3.719~($\pm$0.005)\,\AA. In addition,
using Scherrer formula, we find that the crystallite size (CS) in the
Co-N films is about 6$\pm$1\,nm. A close inspection of this peak
reveals an asymmetry in the peak shape with a shoulder appearing
on the lower 2$\theta$. To get more insight, the XRD data for Co-N films with \pn~=~17.5 to 27.5\p~were normalized and the peak was deconvoluted and fitted assuming two components C1 and C2 as shown in fig.~\ref{fig:xrddcms}(b). Here, C1 correspond to (200) reflection of \tcn~and
C2 to some other Co-N phase as impurity in \tcn. XRD fitting patterns
along with the variations in C1 and C2 are shown in
fig.~\ref{fig:xrddcms}(b). The appearance of a such shoulder can
also be seen in \tcn~films grown in earlier
works~\cite{CoN_AIP_Adv2015,2001_Vac_Asahara,2018_JMMM_NP} and can
be understood due to co-precipitation of a disordered Co$_{3-x}$N
(0$\leq$ x $\leq$1) phase along with \tcn~phase. When the
\pn~increases to 32.5 and 35\p, the presence of Co$_{3-x}$N phase
can be seen even more clearly along with~\tcn~phase.

From the above observations, it can be seen that when \pn~is
between 22.5 to 27.5\p, the variations in the LP are least and the
fraction of the impurity phase is also minimum. This \pn~range
can be most suitable for \tcn~phase formation in the dcMS process. From fig.~\ref{fig:xrddcms}(c) one can also compare the variation of component C1 and LP with \pn. For \pn~=~20 to 25\p~ the area
fraction for C1 is higher which corresponds to \tcn~phase and the
value of LP is close to 3.68\,\AA~(a value observed for
\tcn~phase closer to its theoretical value\cite{CoN_AIP_Adv2015,2019NP_Co_1_3inch}). However, a single
phase \tcn~cannot be realized at any value of \pn. Therefore, it seems that the realization of a single phase \tcn~is
not possible in the dcMS process. In the absence of high~\Ts~and in order to
circumvent this problem, we adopted utilization of HiPIMS
plasma, which is an emerging thin film deposition technique.

\subsection{\tcn~phase formation using HiPIMS}
\label{3.2}

\begin{figure}\center \vspace{-7mm}
    \includegraphics [width=85mm] {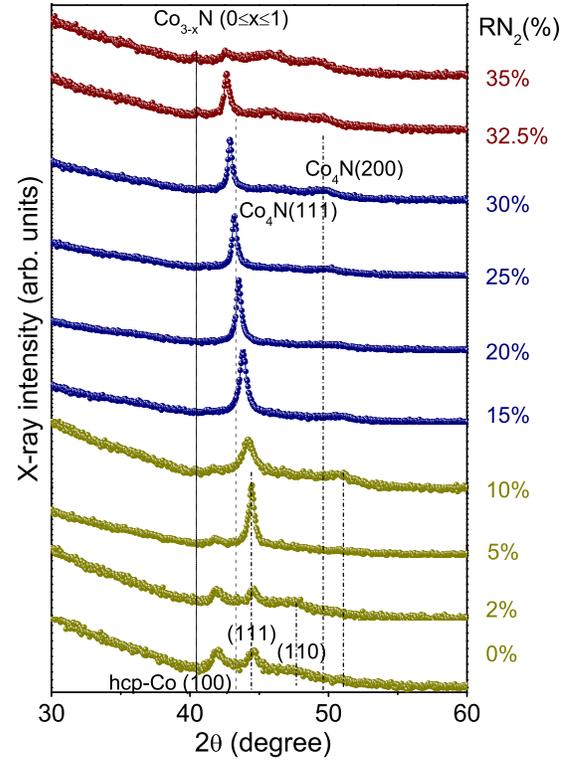} \vspace{-2mm}
    \caption{\label{fig:hipims_lp} XRD patterns of Co-N thin films
        deposited at different \pn~using HiPIMS process.}
\end{figure}

\begin{figure}\center
	\includegraphics [width=85mm] {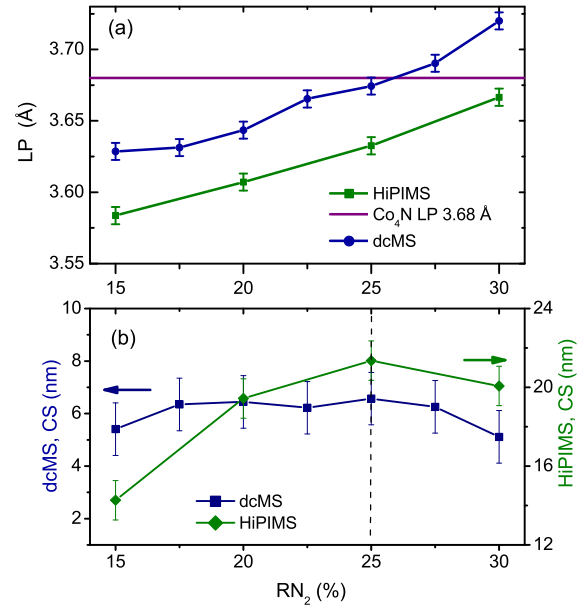}
	\caption{\label{fig:lp_gs} (a) A comparison of lattice parameter (LP) obtained for \tcn~phase deposited using dcMS and HiPIMS process. The horizontal line is a guide to the eye corresponding to LP value of 3.68\,\AA. (b) The variation of crstallite size (CS) as a function of \pn~for Co-N thin films samples grown using dcMS and HiPIMS. It is to be noted that the (200) reflection was used in case of dcMS and the (111), in case of HiPIMS deposited samples.}
\end{figure}

Similar to the dcMS process, Co-N thin films were also deposited at
different \pn using HiPIMS and their XRD
pattern are shown in fig.~\ref{fig:hipims_lp}. As such the evolution of \tcn~phase is similar as that for dcMS process, the striking
differences that can be seen are: (i) the asymmetry observed in the peak profile of samples grown using dcMS is absent in HiPIMS grown samples, (ii) FWHM of the XRD peak is less and (iii) the \tcn~samples adopt
a preferred orientation along (111) direction and the (200) reflection that was dominant in the dcMS grown samples, becomes faint. The absence of asymmetry in the peak profile indicates the impurity phase that invariably was present in samples deposited using dcMS is absent in \tcn~samples grown using HiPIMS. The occurrence of (111) oriented thin films is a general phenomenon that has also been observed for in metal nitrides e.g. TiN, CrN, AlN etc. when grown using HiPIMS process.~\cite{TiN2007,lattemann2010,AlON2019} Using the (111) reflection of \tcn, the LP was obtained and values are plotted in fig.~\ref{fig:lp_gs}(a). As a reference, the theoretical value of LP for \tcn~at 3.68\,\AA~is shown as a horizontal line and values of \tcn~(200) films grown using dcMS process are also included in fig.~\ref{fig:lp_gs}(a) (similar to those shown in fig.~\ref{fig:xrddcms}(c)). It can be seen that the variation of LP with \pn~is quite similar in samples grown using dcMS or HiPIMS, albeit the overall values of LP are somewhat smaller in the later. 

It can also be seen from fig.~\ref{fig:lp_gs}(a) that from \pn~=~15 to 30\p, in films deposited using HiPIMS process, the Co lattice keeps on expanding in a similar manner as that in dcMS process, giving rise to \tcn~phase with
LP~=~3.666($\pm$0.005)\,\AA~for \pn~=~25\p. However, the peak profile is symmetric here and no additional impurity peak can be seen unlike dcMS. A comparison of CS for samples grown using dcMS and HiPIMS is shown in fig.~\ref{fig:lp_gs}(b). It can be seen that the maximum CS for \tcn~phase is nearly 21$\pm$1\,nm, which is significantly higher than \tcn~phase formed by dcMS process at \pn~=~25\p.
For \pn~=~32.5\p~some additional peaks corresponding to N-rich
phases also appear and for \pn~=~35\p~sample, some disordered N-rich phases
starts to form.

The remarkably distinct preferred orientations in Co-N films - (111) in case of HiPIMS and (200) in dcMS, can be understood due to differences in the energetics involved in these two deposition processes. The energy distribution of sputtered atoms in HiPIMS
tails up to 100\,eV~\cite{HIPIMSlundin}. This is reflected in
terms of adatom mobility which is significantly higher in the case
of HiPIMS. Considering the thermodynamics involved as suggested by
Pelleg $et~al.$,~\cite{pelleg1991} the driving force which is
responsible for texture development of thin films consists of
surface ($\gamma$) and strain energy ($\varepsilon$). As
film growth starts at the substrate, there is a competition
between these two processes. It is to be noted here that both
$\gamma$ and $\varepsilon$ have directional dependence for
different $(hkl)$; $\gamma_{100}$ $<$ $\gamma_{111}$ and
$\varepsilon_{111}$ $<$ $\varepsilon_{100}$. In dcMS deposited
films where adatom have low mobility, they tend to have an
arrangement which minimizes surface energy and film orientation is
(200). Higher ion energy involved in the HiPIMS process contribute
to an enhanced adatom surface mobility and the orientation changes to
(111) plane in order to minimize strain energy. Higher adatom mobility also results in larger CS as observed in case of samples deposited using HiPIMS. 

The effect of HiPIMS plasma on morphology of \tcn~phase was also studied in our previous work using atomic force microscopy (AFM) where
well-distributed grains were observed in HiPIMS deposited
\tcn~films while the dcMS deposited \tcn~film does not show
distinguishable surface morphology.\cite{JMMM_MG} The comparison of lattice expansion as a function of \pn~for samples
deposited using dcMS and HiPIMS is shown in
fig.~\ref{fig:lp_gs}(a). Similar kind of expansion has been reported for
Fe-N films deposited using dcMS and HiPIMS~\cite{TAYAL2015}. In
both processes, the trend of expansion of the fcc lattice seems to be
identical. It seems that the highest LP can be obtained for dcMS
deposited film, but the \tcn~(200) reflection was asymmetric and
had additional impurity phase. This
suggests that utilization of HiPIMS plasma favors
\tcn~phase growth as compared to that in the dcMS process. Keeping in mind the
fact that the enthalpy of formation (\dE) for \tcn~$\approx$~0\,eV~\cite{imai2014} and in absence of the substrate
heating, the usage of HiPIMS plasma helps in
growth of a single phase \tcn~ with more than three times larger crystallite size on an average.

\subsection{XAFS studies of Co-N films deposited using dcMS and HiPIMS}\label{3.3}

\begin{figure}\center
    \vspace{1mm}
    \includegraphics [width=85mm,keepaspectratio] {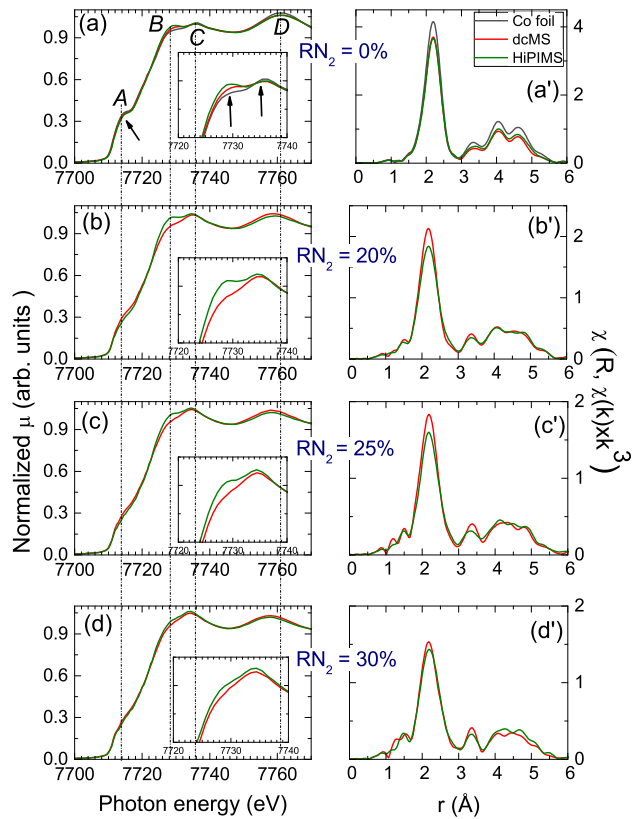}
    \caption{\label{fig:xanes} Co-K edge spectra for Co-N samples deposited at different \pn~using dcMS and HiPIMS. Inset of (a)-(d) show magnified view of feature $B$ and $C$ indicated with arrows. The Fourier transforms of respective absorption spectra is shown in (a$\prime$)-(d$\prime$).} \vspace{1mm}
\end{figure}

From the XRD results presented in previous sections, it can be observed that the insertion of N in Co results in the expansion of the lattice. Moreover, the LP of Co-N thin film samples deposited using dcMS and HiPIMS was found to be less than the theoretically predicted value.~\cite{PRB:CoN:07} To understand this, the local structure of Co-N thin films samples was studied by carrying out x-ray absorption fine structure spectroscopy (XAFS) measurements in nitrided and pure Co samples. Fig.\ref{fig:xanes} (a-d) compares Co K-edge x-ray absorption near edge structure (XANES) spectra of samples deposited at \pn~=~0\p, 20\p, 25\p, and 30\p~with dcMS and HiPIMS processes. The spectra show distinct near-edge features marked as $A$, $B$, $C$, and $D$. These features can be attributed to one-electron excitations and reflect the maximum of p-type density of state of Co which are allowed by the dipole selection rule ($\bigtriangleup l=\pm$~1). The feature $B$ and $D$ are separated by $\approx$~30\,eV, which respectively arises due to surrounding of central Co atom with two nearest neighbors at 2.49 and 2.51\,\AA~\cite{zhang2004xanes}. It can be seen that the intensity of features $B$ and $C$ varies for the samples deposited using different deposition techniques as well as with the change in \pn. As discussed earlier, due to energetic involved in the sputtering process and thermodynamic conditions required for formation of nano-clusters, the formation of different stacking sequences can take place in thin films. It has been observed by Longo $et~al.$~that the higher fraction of the fcc-type stacking sequence gives a higher intensity of feature $B$~\cite{longo2014_Co}. In that report, the fcc-type stacking was attributed to the closed packed sequence. The intensity of feature $B$ found to reduce systematically for the intertwined stacking sequence and show a minimum for the sample having hcp-type stacking sequence. It can be seen that the intensity of the feature $B$ is higher for HiPIMS samples as compared to those grown using dcMS. Therefore, considering theoretical and experimental observations, it can be deduced that a relatively higher fraction of the fcc component is present in the films deposited using HiPIMS process as compared to that in dcMS process. This behavior suggests that the adatom energy, which significantly differs in these two techniques, has a substantial effect in the stabilization of certain crystallographic stacking. In the present case, higher fraction of fcc-Co phase gets stabilized in samples grown using HiPIMS process than that in dcMS.

This trend of higher intensity of feature $B$ was maintained for the sample deposited at \pn~=~20\p, 25\p~and 30\p~suggesting that the fraction of the fcc component is significantly higher for samples prepared using HiPIMS even at higher nitrogen pressure. Since the XRD analysis reveals a simultaneous increment in the lattice parameters with \pn~and fcc-Co and \tcn~share a similar structure, the only difference that the octahedral sites are fully occupied in the \tcn~phase. Therefore, it can be deduced that the films having higher fcc-Co fraction are also likely to have a higher fraction of the fcc \tcn~phase, which may or may not have fully occupied interstitial sites. Similar observation was made by Silva $et~al.$~\cite{2015_Silva}, where XANES spectra of Co K-edge for Co-N film deposited on MgO substrate was studied and it was observed that the structure of \tcn~film is very similar to fcc-Co with a small lattice expansion(0.3\p), indicating a N-deficient or less homogeneity of nitrogen in samples~\cite{2015_Silva}.

Fig.~\ref{fig:xanes}(a$^\prime$-d$^\prime$) shows the FT spectra of indicated samples. It can be seen that for the thin film samples the amplitude of FT peaks are smaller compared to the bulk Co (Fig.~\ref{fig:xanes}(a$^\prime$)). Such reduction in the FT amplitude can be attributed to the smaller particle size or disorders in the nanosized samples~\cite{Frenkel:2011,PRB:Frenkel}. Interestingly, the amplitude of first FT peak is slightly smaller for the HiPIMS samples relative to the dcMS as shown in fig.~\ref{fig:xanes}(a$^\prime$). With increasing \pn, the first peak amplitude remains relatively smaller for the HiPIMS samples (fig.~\ref{fig:xanes}(b$^\prime$-d$^\prime$)). However, the FT amplitude in the region between 4--5~\AA~is comparable in both samples. In this region, the forward scattering shells dominates the FT spectrum~\cite{Teo:1981} whose contribution is significantly higher for the fcc structure. It suggests that the amplitude ratio of the MS shells with the first FT peak is higher for the samples deposited using HiPIMS as compare to the dcMS indicating a significant fraction of the fcc component in these samples. A similar trend for the relative amplitude of the MS peak and the first peak is observed for the samples deposited up to \pn~=~30\p. 

To get further information on the evolution of local structure in Co-N films, FT spectra for hcp-Co, fcc-Co and \tcn~were simulated as shown in fig. S1 in the SM.~\cite{SM} Since the FT spectrum can be correlated to the separate atomic shells, subtle variation in the local structure of hcp-Co, fcc-Co and \tcn~can be used to obtain qualitative information on the relative fraction of these phases present in the sample. The first FT peak around 2.2~\AA~is associated with the first shell at 2.5~\AA~in the fcc structure. This peak in the hcp structure comes from the contribution of two closely related shells (2.48~\AA~and 2.50~\AA~), which produces a single peak in the FT spectrum. The peaks at around 3.3\,\AA~ and 4\,\AA~are contributed from the second shell at 3.52\,\AA~and third shell at 4.3\,\AA, respectively. The region between 4.2--5\,\AA~is dominated by multiple scattering (MS) shells. In the fcc structure, due to the presence of collinear chains of three atom at the double the distance of the first shell, a contribution to the FT amplitude form such MS shells is much higher compared to the single shell. Therefore, the FT amplitude is higher in this region for the fcc structure, as demonstrated in the simulated FT spectra on hcp and fcc Co (see fig.S1 in SM~\cite{SM}). 

 Consequently, a relative amplitude of higher-order MS shell with the first FT peak can act as a useful way to qualitatively estimate the relative fraction of the fcc component in the thin films.  It can be seen that for the pure Co (\pn~=0\p), the intensity of the first FT peak is nearly identical for the samples deposited using dcMS and HiPIMS. However, the FT amplitude in the region between 4-5~\AA~is slightly smaller for the sample prepared with dcMS as compared to HiPIMS. It suggests that for the pure Co film (\pn~=~0\p) prepared using HiPIMS having a higher fraction of the fcc component relative to the dcMS, as also observed in the XANES analysis discussed previously. For the fcc \tcn, the FT peak shifts to a higher value due to increase in the lattice parameter (see fig. S1 in SM~\cite{SM}). The region between 4--5~\AA~still remains dominated by the forward scattering MS shells in this structure. Interestingly, the FT peak at around 3.2~\AA~has a similar intensity for the fcc or hcp Co. However, it has a significantly lower amplitude for the \tcn~phase as shown in fig. S1 of SM.~\cite{SM} For the Co-N samples it can be seen that the FT amplitude in this region starts to decrease from 0\p~to 20\p~\pn~and show minimum value at 25\p~for the samples prepared using HiPIMS. Moreover, the FT amplitude is smaller for the HiPIMS samples compared to the dcMS. It indicates that the relative fraction of \tcn~phase is higher in the HiPIMS sample and at \pn~=~25\p~it gets maximized. 

\subsection{EXAFS analysis of Co-N films deposited using dcMS and HiPIMS}\label{3.4}

\begin{figure*}\center
	\vspace{1mm}
	\includegraphics [width=110mm,keepaspectratio] {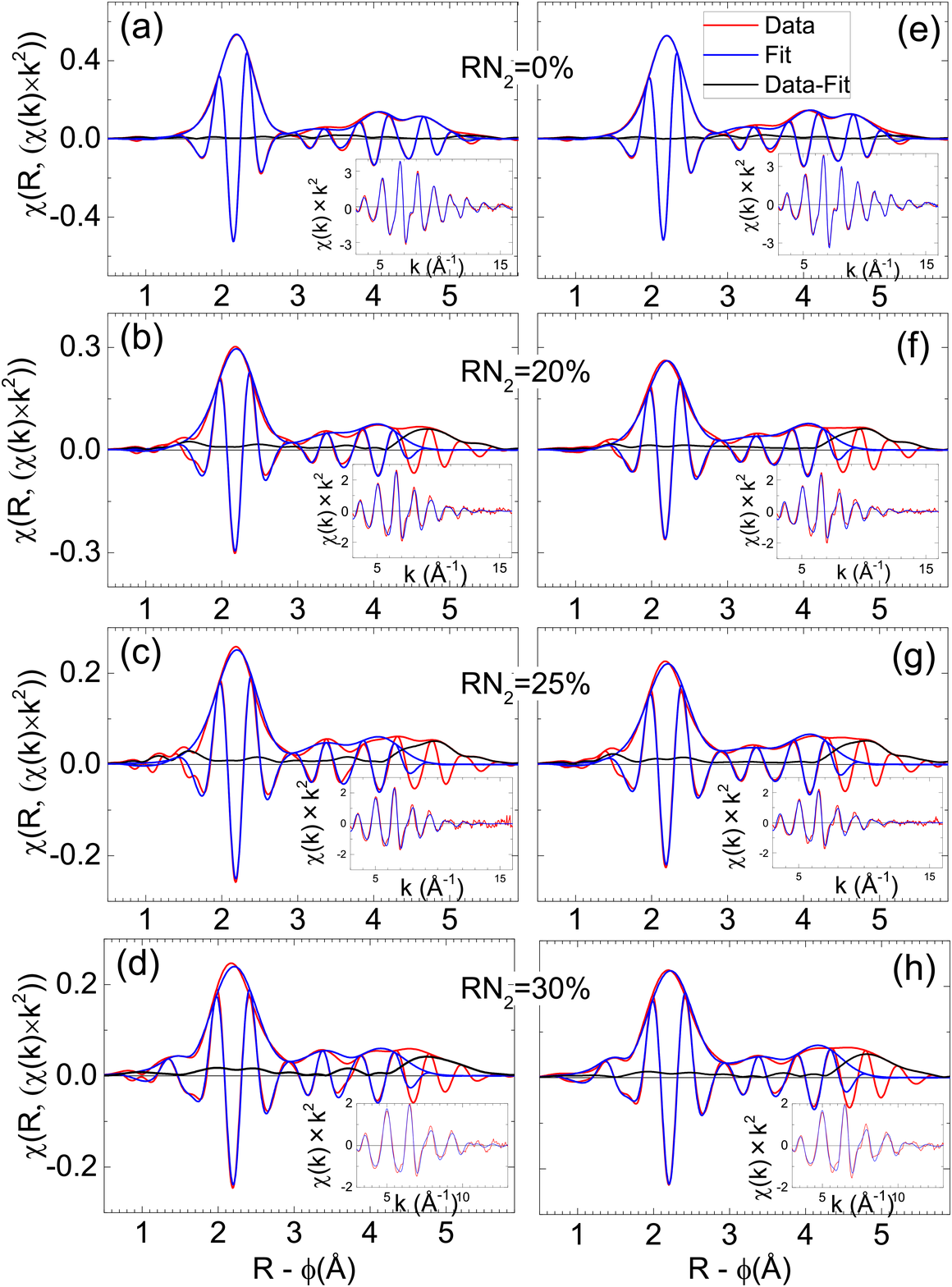}
	\caption{\label{fig:exafs1} Fourier transform moduli of the Co $K$-edge EXAFS ($|\chi(R)|$), and real component (Re$[\chi(R)]$) of the data and fit of dcMS(a,b,c,d) and HiPIMS(e,f,g,h) samples at different \pn. Inset shows $\chi (k) \times k^2$ data and fit. The x-scale ($R-\phi$) is phase-shift uncorrected atomic pair distance.}
\end{figure*}

\begin{table*} \centering
	\caption{\label{tab:EXAFS}  EXAFS metrical parameters derived from the fittings for selective Co-N thin films measured at Co $K$-edge. Here, $R$ is obtained atomic pair distance, $N$ is
		the coordination number, $\sigma$ is the root mean square displacement and
		$\Delta E_0$ energy is shift parameter. The parameters reported
		without error bars were kept fixed during the fitting. The value of $\Delta E_0$ for all the shells was kept constrained to vary by $\pm$1\,eV relative to the first shell.}
	\footnotesize
	\begin{tabular}{|c|c|c|c|c|c|c|c|c|c}
		\hline\hline \textbf{dcMS} &	Shell	&	I	&	II	&	III	&	IV	&	V	&	MS	&	MS \\
		\hline
		&	Pair	&	Co-Co	&	Co-Co	&	Co-Co	&	Co-Co	&	Co-N	&	MS	&	MS \\
		\hline
		\pn~=~0\p	&	$R$ (\AA)          &2.48$\pm$0.01        &3.51$\pm$0.02        &4.34$\pm$0.02     &&&4.95$\pm$0.02 & \\
		&	$N$                  &10.4$\pm$0.1          &4.2$\pm$0.1          &14.8$\pm$4.3           &&&5.8$\pm$1.5    & \\
		&	$\sigma$           &0.07$\pm$0.01        &0.09$\pm$0.01        &0.09$\pm$0.01        &&&0.09$\pm$0.01  & \\
		&	$\Delta E_0$         &4.8$\pm$1.9        &           &          &&& &\\
		\hline
		
		\pn~=~20\p	&	$R$ (\AA)          &2.50$\pm$0.01        &3.55$\pm$0.02        &3.73$\pm$0.02     &4.39$\pm$0.02         &&4.99$\pm$0.02&5.25$\pm$0.02 \\
		&	$N$                  &10.8$\pm$2.3          &5.0$\pm$1.5          &6.0$\pm$1.6           &18.0$\pm$5.1            &&2.0$\pm$0.6&8.4$\pm$2.5\\
		&	$\sigma$           &0.10$\pm$0.01        &0.11$\pm$0.01        &0.15$\pm$0.01        &0.11$\pm$0.02  &&0.11& 0.11\\
		&	$\Delta E_0$         &2.5$\pm$2.2          &           &           &      &&3.00\\
		\hline
		
		\pn~=~25\p	&	$R$ (\AA)          &2.51$\pm$0.01        &3.53$\pm$0.02        &3.73$\pm$0.03     &4.40$\pm$0.02         &&5.00$\pm$0.02& 5.17$\pm$0.02\\
		&	$N$                  &11.6$\pm$2.6          &3.9$\pm$1.3          &5.9$\pm$1.5           &17.8$\pm$5.1            &&1.7$\pm$0.5&7.2$\pm$2.1\\
		&	$\sigma$           &0.11$\pm$0.01        &0.10$\pm$0.01        &0.17$\pm$0.01        &0.12$\pm$0.02  &&0.11&0.11 \\
		&	$\Delta E_0$         &2.5$\pm$2.4           &           &          &     &&3.00\\
		\hline
		
		\pn~=~30\p	&	$R$ (\AA)          &2.53$\pm$0.01        &3.58$\pm$0.02        &3.72$\pm$0.02     &4.45$\pm$0.02         &1.85$\pm$0.02&5.03$\pm$0.02& 5.21$\pm$0.02\\
		&	$N$                  &11.0$\pm$2.2          &4.9$\pm$1.3          &3.6$\pm$1.1           &17.8$\pm$5.3            &0.8$\pm$0.2&2.1$\pm$0.6&4.4$\pm$1.3\\
		&	$\sigma$           &0.11$\pm$0.01        &0.11$\pm$0.01        &0.11$\pm$0.01        &0.13$\pm$0.01  &0.06$\pm$0.03&0.11&0.11 \\
		&	$\Delta E_0$         &3.3$\pm$2.2           &           &           &      &&3.00&\\
		\hline\hline
		
		\textbf{HiPIMS} &~Shell~&~I~&~II~&~III~&~IV&~V~&~MS~&~MS~& \\
		\hline
		\pn~=~0\p	&	$R$ (\AA)          &2.48$\pm$0.01       &3.51$\pm$0.02        &4.34$\pm$0.01     &&&4.95$\pm$0.01   & \\
		&	$N$                  &10.2$\pm$1.7          &4.6$\pm$1.1          &16.5$\pm$4.5           &&&6.8$\pm$1.7        &\\
		&	$\sigma$           &0.07$\pm$0.01        &0.09$\pm$0.01        &0.09$\pm$0.01        &&&0.09$\pm$0.01  &\\
		&	$\Delta E_0$      &5.2$\pm$2.0           &           &           &&&  &\\
		\hline
		\pn~=~20\p	&	$R$ (\AA)   &2.51$\pm$0.01        &3.52$\pm$0.02        &3.84$\pm$0.03     &4.39$\pm$0.02         &&5.03$\pm$0.02& 5.12$\pm$0.02 &\\
		&	$N$                  &9.9$\pm$2.2          &4.9$\pm$1.6          &6.9$\pm$1.7           &21.7$\pm$6.1            &&0.8$\pm$0.2	&	10.7$\pm$3.0 \\
		&	$\sigma$           &0.10$\pm$0.01        &0.12$\pm$0.01        &0.19$\pm$0.01        &0.12$\pm$0.01  &&	0.11	&	0.11	& \\
		&	$\Delta E_0$         &3.4$\pm$2.5          &           &          &      &&3.00	&\\
		\hline
		\pn~=~25\p	&	$R$ (\AA)   &2.52$\pm$0.01        &3.51$\pm$0.02        &3.87$\pm$0.04     &4.41$\pm$0.02         &&5.04$\pm$0.02& 5.16$\pm$0.02 &\\
		&	$N$                  &10.5$\pm$2.3          &4.8$\pm$1.5          &6.8$\pm$1.5           &21.1$\pm$5.9            &&0.7$\pm$0.2	&	9.3$\pm$2.6 \\
		&	$\sigma$           &0.11$\pm$0.01        &0.12$\pm$0.01        &0.20$\pm$0.01        &0.12$\pm$0.01  &&	0.11	&	0.11 \\
		&	$\Delta E_0$         &3.0$\pm$2.5         &           &           &      &&3.00	&\\
		\hline
		\pn~=~30\p	&	$R$ (\AA)          &2.54$\pm$0.01        &3.58$\pm$0.02        &3.74$\pm$0.03     &4.47$\pm$0.02         &1.86$\pm$0.02&5.03$\pm$0.02& 5.21$\pm$0.02\\
		&	$N$                  &9.9$\pm$2.1          &5.2$\pm$1.4          &2.8$\pm$0.8           &18.6$\pm$5.4            &0.8$\pm$0.2&	2.1$\pm$0.6&	4.4$\pm$1.3\\
		&	$\sigma$           &0.11$\pm$0.01        &0.12$\pm$0.01        &0.11$\pm$0.01        &0.12$\pm$0.01  &	0.05$\pm$0.042&	0.11&	0.11 \\
		&	$\Delta E_0$         &3.4$\pm$2.2           &          &           &      &&	3.00&\\
		\hline\hline
		
	\end{tabular}
	\vspace{-2mm}
\end{table*}

To gain a further insight on the local structure of Co-N films, detailed EXAFS fitting was performed to estimate the atomic pair distance ($R$), coordination number ($N$), and root mean square displacement ($\sigma$). The value of passive electron reduction factor (0.8) was empirically obtained from fitting bulk Co and kept fixed for all of the samples. As discussed earlier, the local structure of fcc-Co, hcp-Co and \tcn~differs only slightly. The fcc-Co and hcp-Co have identical local structure, however the contribution from the forward scattering MS shells is substantial for the fcc structure. This is an indicator to make a distinction in their structure. The \tcn~and fcc-Co, however, have similar local structure with only higher LP for the former and therefore, the atomic shells are relatively expanded. Therefore, it becomes very challenging to distinguish between the local structure of off-stoichiometric \tcn~phase with fcc-Co. This problem gets further elevated for the heterogeneous samples. Keeping all these things in mind, the EXAFS analysis was performed by fitting the spectra with the single scattering atomic shells which are similar for the hcp, fcc and \tcn~phases. The fitted data was later subtracted from the $\chi(k)$ data to filter out the contribution of remaining MS shells. The filtered data was fitted with the MS shell observed in the fcc-Co and \tcn~structures. A relative variation in the amplitude of MS shells was used to qualitatively estimate the fraction of different phases present in the dcMS and HiPIMS samples at different \pn. 

For the \pn~= 0\p~sample, it would either have a hcp or fcc Co phase. Therefore, the EXAFS fitting for the \pn~=~0\p~dcMS and HiPIMS samples were performed using four shells as shown in fig.~\ref{fig:exafs1}. The first three are the single scattering shells and the fourth is the MS shell. The obtained fitting values are tabulated in the table~\ref{tab:EXAFS}. It can be seen from this table that metrical parameters for all the shells are similar except the $N$ for the forward scattering MS shell. It was found that for the HiPIMS sample it has a slightly higher value indicating relatively greater fraction of the fcc component in the sample. It must be noted that the region between 4--5\,\AA~have contribution from the forward as well as double forward scattering shells with a small fraction of single scattering shell. Therefore, the $N$ and $\sigma$ is contributed from all these shells and the obtained values may have some error relative to the absolute value. In order to avoid larger number of independent fitting parameters, these were kept free in the fitting procedure and only forward scattering shell was used. This assumption do not go against the motivation of this study which is to obtain only relative change in the $N$. Moreover, the metrical parameters give quantitative estimation which are also clearly evident in the raw EXAFS spectra. 
 
The XRD analysis of dcMS deposited samples reveals that at \pn~=~20\p~and 25\p~the formation of either off-stoichiometric \tcn~phase or mixture with hcp-Co phase occurs. However, in HiPIMS deposited samples no signatures corresponding to the hcp-Co could be seen, means its contribution if present, would be below the background level. Therefore, to distinguish between these phases the contribution from the MS shells was filtered from the $\chi(k)$ data which was initially fitted with four shells and the obtained values are listed in table~\ref{tab:EXAFS}. In addition to three single scattering shells, one MS (Co-N-Co) was only included due to forward scattering geometry. It can be seen that with increase in \pn, the atomic pair distance also increases and the HiPIMS sample has slightly smaller atomic pair distance relative to dcMS for the second shell which corroborates with the XRD results. 

To precisely determine the phase fraction of \tcn, hcp-Co and fcc-Co in the dcMS and HiPIMS deposited samples, the residual $\chi(k)$ data was fitted with MS shells of fcc-Co and \tcn~structure as shown in fig. S2 (see SM~\cite{SM}). The fig. S3 in SM shows the subtracted $\chi(k)$ spectra of indicated samples with different \pn. It can be seen that for \pn~=~20\p, the FT peak is slightly shifted to a higher value for HiPIMS sample. Similar trend can be seen for the \pn~=~25\p~sample. Such a shift in the FT peak position towards a higher value is reflected in the EXAFS fitting. It was observed that for the HiPIMS sample, the inclusion of MS shell corresponding to the \tcn~phase could only fit the data reasonably. Moreover, the value of coordination number for this shell successively increases with \pn~for the HiPMS sample. This observation indicates that the presence of \tcn~phase is relatively higher in the HiPIMS samples which is clearly an effect of dense plasma formed during the deposition process. Above 25\p, the formation of N-rich Co$_3$N phase occurs as seen in the loss in the FT amplitude for the 30\p~sample relative to 25\p~(see fig. S4 in SM~\cite{SM}). Due to different coordination geometry the FT amplitude is smaller for the atomic pairs in the Co$_3$N structure. The EXAFS fitting of 30\p~sample was performed following the similar process as discussed previously (see Table~\ref{tab:EXAFS} and Fig.~\ref{fig:exafs1}). It was found that inclusion of Co-N shell improves the quality of the fit and contribution from the forward scattering MS shell reduces substantially for the 30\p~sample(Fig. S2 and S3 in SM~\cite{SM}) indicating formation of nitrogen rich Co$_3$N phase. The presence of Co$_3$N phase along with \tcn~can deteriorate the quality of the sample. Therefore, at \pn~=~25\p~the \tcn~phase fraction was found to be maximum from XRD as well as EXAFS analysis.

\subsection{Magnetic properties of Co-N films}\label{3.5}
\begin{figure}\center
	\vspace{-1mm}
	\includegraphics [width=100mm] {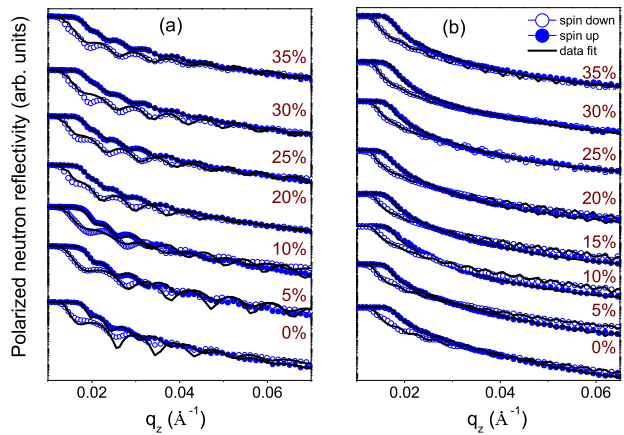} 	\vspace{-5mm}
	\caption{\label{fig:pnrall} Polarized neutron reflectivity (PNR) patterns of Co-N films deposited at different \pn~values using dcMS (a) and HiPIMS (b).} \vspace{-1mm}
\end{figure}

\begin{figure}\center
	\vspace{-1mm}
	\includegraphics [width=85mm] {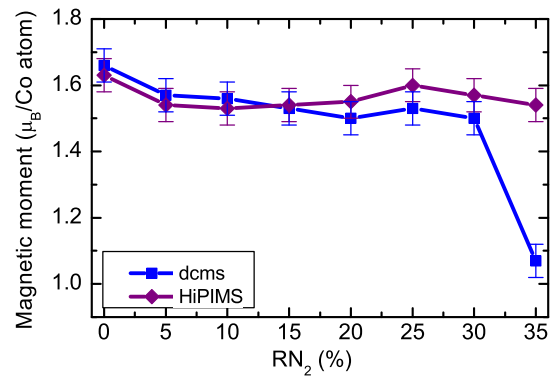} 	\vspace{-2mm}
	\caption{\label{fig:moment} Magnetic moment of Co-N thin films deposited using dcMS and HiPIMS process at different \pn~values obtained from fitting of PNR data.} \vspace{-1mm}
\end{figure}

Since Co based nitride compounds are well-known for their ferromagnetic properties, the magnetization of Co-N films have been studied combining different techniques such as S-VSM, PNR and MOKE. For thin films, PNR is the accurate method
to estimate the magnetic moment (\ms) as errors arising due to sample volume
determination and the diamagnetic contribution from the
substrate are not present. The PNR data of Co-N samples prepared
at different \pn~using dcMS and
HiPIMS is shown in fig.~\ref{fig:pnrall}(a,b). The mass
densities obtained from the XRR data fitting (not shown) were
used to fit the nuclear scattering length density profile of PNR
at 300\,K.~\cite{JMMM_MG} All samples show ferromagnetic behavior, as expected
with clear splitting between spin-up (R+) and down (R-)
reflectivity. The obtained values of magnetic moment (\muB) per Co
atom is plotted against \pn~as shown in fig.~\ref{fig:moment}.

For pure Co films prepared by dcMS or HiPIMS, the value of the moment is 1.60\muB($\pm$0.05) which
matches well with theoretical and experimental values in
literature.~\cite{PRB:CoN:07,zhao2016metastable} As \pn~increased, the moment reduces
as the hcp structure starts to distort by N incorporation in the
Co lattice (see fig.~\ref{fig:xrddcms}). From \pn~=~10\p~to
20\p~the value of moment remains almost constant (within the error
bar) as the lattice becomes fcc due to N incorporation. This
happens due to a magneto-volume effect which leads to high
volume-high moment phenomenon at corner sites of
\tcn~lattice~\cite{PRB:CoN:07}. After \pn~=~20\p~the \ms~increases
which indicates~\tcn~phase formation as also observed in XRD measurements.
However, for the \pn~range of 20-30\p~HiPIMS deposited samples show
slightly higher moment as compared to dcMS, which may happen as
the films deposited using HiPIMS have larger crystallite size, contains no additional impurity phase unlike dcMS and higher \tcn~phase fraction, as observed by XRD and XAFS analysis.

\begin{figure}
    \includegraphics [width=85mm] {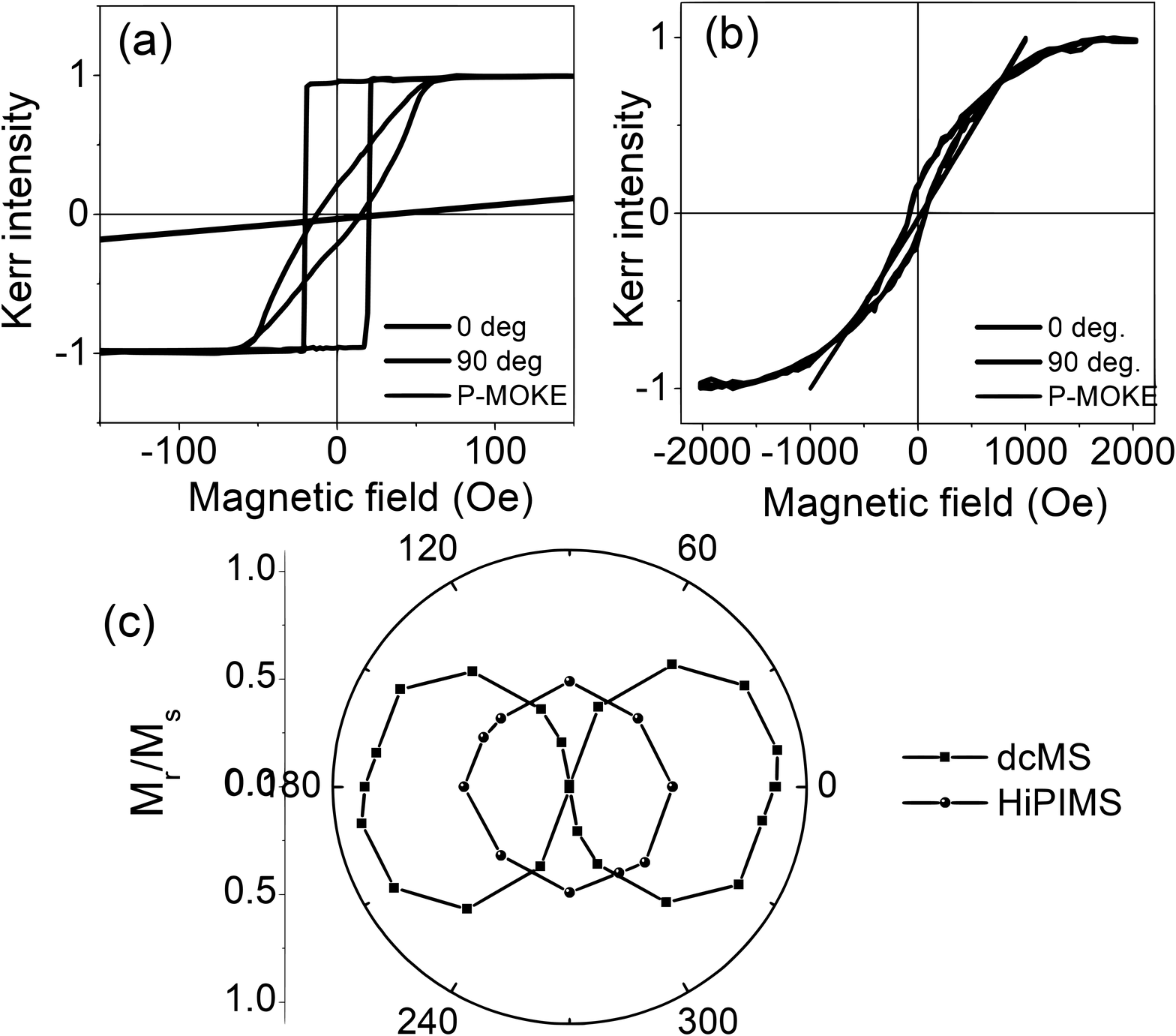}
    \caption{\label{fig:polar} Kerr intensity variation with magnetic
        field in longitudinal (L) and polar (P) geometry for \tcn~samples deposited at \pn~=~25\p~prepared using (a) dcMS and (b) HiPIMS.
        Angular variation of squareness (Mr/Ms) for these samples (c).}
    \vspace{-1mm}
\end{figure}

To further investigate magnetic properties of \tcn~films, the magnetic hysteresis (M-H) data was collected by L-MOKE and P-MOKE
measurements on \tcn~samples (\pn~=~25\p). M-H loops were taken by
keeping the laser spot position fixed, but rotating the sample to
investigate the in-plane magnetic anisotropy. Samples were rotated
from 0 to 180$^{\circ}$ in steps of 20-30$^{\circ}$ with the field
applied parallel to the film in L-MOKE set up.
Fig.~\ref{fig:polar} shows M-H loops taken for \tcn~films
deposited using dcMS and HiPIMS. It can be seen that for dcMS deposited \tcn~films, there
exist separate easy and hard axis of magnetization which is generally not observed in polycrystalline films. However, for
HiPIMS deposited \tcn~film there is no easy axis of magnetization
and also the field required to saturate the film is quite high. No
signatures of perpendicular component of anisotropy were found in
any \tcn~sample from P-MOKE measurements. Squareness (Mr/Ms) and
in plane rotation angle ($\phi$) have been plotted and shown in
fig.~\ref{fig:polar}~(c). From the variation of Mr/Ms with the
in-plane rotation angle, the two-fold symmetry axis was found in
dcMS deposited films indicating in-plane uniaxial anisotropy. This might happen due to the additional phase fraction (hcp-Co) in form of nano-clusters formed during dcMS process.
However, for film deposited using HiPIMS, Mr/Ms do not reveal any
directional dependence within the film plane, implying that the
film is magnetically isotropic in nature, which is generally expected for a polycrystalline film. As we already discussed in XAFS analysis, the major phase in HiPIMS deposited film is fcc-\tcn~along with a minor fcc-Co. On the other hand, the \ms~value for \pn~=~25\p~samples was found to be almost similar by PNR measurements. This indicates that growth process has
a significant effect on magnetic properties of \tcn~phase and higher fcc phase fraction present in HiPIMS deposited films lead to
magnetically isotropic nature of these films.

\begin{figure}\center
    \includegraphics [width=85mm] {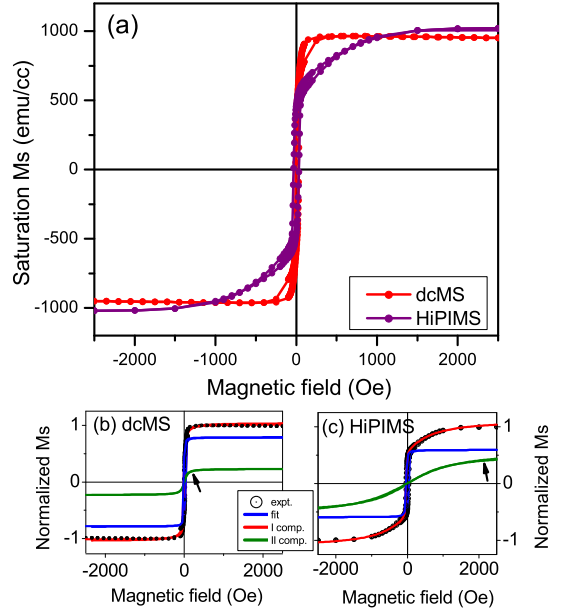}
    \caption{\label{fig:svsm} (a) SVSM data of \tcn~samples deposited
        using dcMS and HiPIMS. M-H plots fitted with 
        components I and II (as described in eq.~\ref{E}) for \tcn~sample
        deposited using dcMS (b) and HiPIMS
        (c).} \vspace{-4mm}
\end{figure}

To elucidate the in-plane magnetization in detail,
room-temperature S-VSM measurements were carried out on
\tcn~samples (\pn~=~25\p) and obtained loops are shown in
fig.~\ref{fig:svsm}(a). A closer inspection of M-H loop shows
characteristic features of two-phase system, as the positive and
negative field cycle shows different saturation behavior. This is
confirmed by fitting the normalized M-H loops using two components
(labelled as I and II in fig.~\ref{fig:svsm}(b,c)) in the
following equation~\cite{PRB:AT:2014}:
\begin{equation}\label{E}
M(H)=\sum\limits_{i=1}^{n}\frac{2M_s^i}{\pi}\arctan\left|\left(\frac{H\pm~H_c^i}{H_c}\right)\tan\left(\frac{\pi~S^i}{2}\right)\right|
\end{equation}
here, M$_s^i$ is the saturation magnetization, H$_C^i$ is the
coercivity and S$^i$ is the ratio Mr/Ms of $i$$^\mathrm{th}$ component of M-H
loop. Here, the component I correspond to \tcn~phase and gets
saturated at very low applied field ($\approx$ 200\,Oe). On the
other hand, the component II may originate due some additional
magnetic phase. This component II has a similar appearance in dcMS
grown samples (fig.~\ref{fig:svsm}(b,c)) but in HiPIMS grown
sample, it shows a completely different behavior. As can be seen
from fig.~\ref{fig:svsm}(c), the component II does not get
saturated even up to $\approx$~2000\,Oe (indicated with arrow in fig.\ref{fig:svsm}). Therefore, this component
which gets saturated at high field value does not show any
contribution in anisotropy leading magnetically isotropic
\tcn~film. These findings are also in accordance with the previous obtained results in XRD, XANES and MOKE measurements. In
HiPIMS, higher ionized plasma density as compared
to dcMS process enhances reaction of N and Co. In HiPIMS process,
energetic adatom as well as higher degree of ionized plasma have led to
formation of fcc-Co (II component) along with \tcn~which was hcp-Co in films deposited using dcMS and observed in local structure analysis by XANES.

\section*{Conclusion}
\label{4}In this work, we systematically studied the phase formation of \tcn~utilizing dcMS and HiPIMS processes. The growth process, structure, electronic and magnetic properties have been studied. It was observed that
the difference in growth mechanism due to higher adatom mobility in the HiPIMS process resulted in (111) oriented Co-N films on the amorphous quartz substrate. XAFS analysis on Co-N films shows formation of an additional phase in the Co-N films along with \tcn~phase. This phase was attributed to be in the form of nano-clusters consisting of mixtures of hcp and fcc Co. The phase fraction of \tcn~was found to high in HiPIMS deposited films as compared to dcMS. Saturation magnetization of all \tcn~films was found to be almost similar from PNR and S-VSM measurements. From MOKE measurement, the HiPIMS deposited \tcn~films were found to be magnetically isotropic. Presence of an additional ferromagnetic component (fcc-Co) in
HiPIMS deposited \tcn~film was observed in M-H data which have lead to loss of magnetic anisotropy of the film.

\section*{Acknowledgments}
This project has received funding from the EU-H2020 research and
innovation programme under grant agreement no. 654360 having
benefitted from the access provided by JCNS, Garching in Germany
within the framework of the NFFA-Europe Transnational Access
Activity. Department of Science and Technology, India
(SR/NM/Z-07/2015) is acknowledged for the financial support and
Jawaharlal Nehru Centre for Advanced Scientific Research (JNCASR)
for managing the project. We would like to thank L. Behera for the
help provided in various measurements. We are also thankful to V.
R. Reddy, A. Gome, M. Kuila for XRR and MOKE measurements, S. Majumder and R. J. Chaudhary for S-VSM measurements. We are thankful to A. Banerjee and A. K. Sinha for support and
encouragement.


%

\end{document}